\documentstyle[epsf,preprint,floats,tighten,aps]{revtex}
\begin{document}
\title{Refined geometry and frozen phonons in KNbO$_3$}
\author{A.~V.~Postnikov\cite{*} and G.~Borstel}
\address{Department of Physics, University of Osnabr\"uck,
	 D-49069 Osnabr\"uck, Germany}
\maketitle
\begin{abstract}
Full-potential LMTO calculations for KNbO$_3$ reported up to now
provided a reasonable description of off-center equilibrium
displacements, including transversal optical $\Gamma$ phonons.
However, when addressing more sensitive phonon properties
and the behavior of the ferroelectric instability
over the Brillouin zone, the need was realized
to achieve a substantially higher level of accuracy.

In order to arrive at ultimately accurate results
available with the LMTO method in the local density approximation,
the stability of full-potential LMTO predictions
for off-center displacements in KNbO$_3$,
as depending on the choice of basis and expansion cutoffs,
has been thoroughly investigated.

With the calculation setup thus optimized,
supercell frozen phonon calculations
aimed at the study of the chain-structure
instability over the Brillouin zone
have been done, and the long-wavelength limit
of the LO phonon is discussed.
\end{abstract}

\section{Introduction}
\label{sec:intro}

In referring to the nature of the ferroelectric instability
in KNbO$_3$, which develops as Nb atoms are displaced
from their symmetric positions in the centre of the cubic perovskite cell,
it became usual to mention a crossover from displacive
to order-disorder type of transition, that gave rise
to considerable controversy \cite{fmsg84,frkc88,scr88,fkc89,scr89}.
Indeed, both the temperature dependence
of the dielectric constant \cite{js62}
and the anisotropy of the transverse acoustic and lowest optical
phonon branches may be understood in the most natural way
in the framework of the displacive model \cite{fkmc81}.
At the same time, recent stimulated Raman scattering
experiments \cite{dwng92,doug1,doug2}
clearly suggest that separate off-center potential minima
for Nb atoms exist, and the hoppings between them
can be induced by an appropriate photon pumping.  This controversy
may be understood in the following way: The displacive scenario
supposes that the (in our case Nb) atoms are sitting at,
or rather oscillating about, the central symmetric
position in the high-temperature paraelectric phase,
and this is definitely not the case in KNbO$_3$. In fact
Nb atoms are displaced in each individual unit cell
all way around through the sequence of phase transitions.
But the textbook description of the order-disorder model
with Nb atoms statistically distributed over eight
[111]-displaced positions is also not true in its implicit
assumption that each individual
atom has a choice of eight local potential minima
around the centre of the perovskite cell.
Be it so, the scan of the potential
hypersurface related to the displacement of a {\em single}
individual atom in an otherwise non-polar paraelectric
crystal would clearly exhibit these minima. Such scanning,
that is difficult to implement experimentally but easy to imitate
in a supercell total-energy calculation,
shows no traces of off-center minima and
no noticeable anharmonic damping of local potential surface
related to a {\em single} atom displacement
beyond any uncertainty of calculation\cite{wil94}.
At the same time, precise total-energy calculations
for perfectly ordered ferroelectric KNbO$_3$\cite{sb92,ktn3,ksv94}
routinely reproduce the tendency of Nb atoms
to displace {\em simultaneously} off-center
in the [111]-direction, and the $\Gamma$-TO frequencies
which come out reasonably well in such calculations
\cite{sb92,phonon,ortho,loto} suggest that the
potential hypersurface related to various atomic displacements
is described sufficiently well.

The message from this mixed evidence is that
the off-center potential minimum for the Nb atom in each
individual perovskite cell is determined by the electric field
in some vicinity of the cell in question. In other words,
the displacement of Nb atoms is correlated,
and the ferroelectric instability develops as the softening
of particular phonon modes. This statement belongs
to the vocabulary of the displacive model. The important
difference is that the particular vibration pattern may have
finite spatial extent, or correlation length, even at
very low temperatures. Another difference is that
a particular soft mode may not necessarily freeze down
with temperature, but rather remain frozen at all temperatures
throughout the ferroelectric transitions.
What evolves with temperature is either the statistical mixture,
or the dynamical interplay, of different frozen modes,
apparently accompanied by a gradual decrease
of the correlation length.

With respect to spatially correlated displaced Nb atoms,
a guess that they may build chain-like structures
has been put forward long ago by Comes {\em et al.} \cite{comes}
and since then addressed by new measurements
(see, e.g., Ref. \cite{chen}).
It was not however possible to extract experimentally
any information as for the correlation length
along the chain, nor to whether there is any
correlation between different chains.
{}From the lattice-dynamical point of view, the study
of the spatial correlation along the chain has to do
with the instability region, i.e. that of the
soft-mode freezing, in some vicinity of the Brillouin zone
center. Yu and Krakauer \cite{chain} mapped the phonon
spectrum of cubic KNbO$_3$ over the whole Brillouin zone
in an {\em ab initio} calculation using a linear response
approach. Once the general shape of the instability region
is known, additional information on the correlation
in the atomic displacements, which may as well include
the effects beyond the harmonic approximation, may be
obtained from supercell calculations. In the present paper,
we provide a preliminary attempt of such analysis,
based on the use of a fast full-potential
linear muffin-tin orbital (LMTO) method.

\section{Refined calculation setup
	 for the full-potential LMTO calculations
	 of KN\lowercase{b}O$_3$}
\label{sec:setup}

Using the fast and accurate full-potential LMTO code
by M.Methfessel\cite{msm1,msm2}
as a tool for {\em ab initio} total energy calculations,
we were able to study earlier in a series of
papers \cite{wil94,ktn3,phonon,ortho}
different aspects of ferroelectricity in KNbO$_3$,
including magnitudes of equilibrium off-center
atomic displacements, coupling of the displacements
to the (tetragonal) lattice strain, and $\Gamma$-TO phonon
frequencies in different phases of this compound.
Keeping in mind the planned applications to large supercells,
we tried to analyze the resources of the method and of our
previously used setup concerning an eventual decrease
in the computational demand without substantial loss
of accuracy. At the same time, we tried to find a way
to improve the description of the O--K bonding,
which came out somehow too rigid in our previous calculations.
As a result, the coupling of K to the oxygen cage
on a ferroelectric transition was overestimated --
see the discussion in Ref.\cite{phonon}.

We considered the refinement of our calculation setup
along the following guidelines:
The choice of the muffin-tin spheres sizes is based on the
spatial distribution of the potential over the cell
(with the exception of K whose sphere is enlarged
in order to include some interstitial space around it)
and need not to be changed compared to as discussed
in Ref. \cite{ktn3}; the same applies to the kinetic energies
of the Hankel function envelopes ($-$0.01, $-$1.0 and $-$2.3 Ry)
which provide sufficient variational freedom.
As has been discussed at length in Section III of Ref. \cite{ktn3},
one possible source of uncertainty in the calculation results is
the termination of the spherical harmonic expansions
of the charge density in the interstitial region at some value
$L_{max}$ (in principle different for each inequivalent atom type).
The usual way to control this is to check the convergency of results
depending on this parameter. Another option in the adjustment of
the calculation setup is the choice of the basis LMTO's.
As a reference point representing convergency with respect
to both these factors, we considered the calculation
with $L_{max}$=6 on all atoms, and all valence-state
basis functions ($s$, $p$ and $d$) included in the basis
set for each value of the Hankel function energy.
The systematic analysis of the trends in the calculated
properties (we concentrated mostly on the magnitude
of the equilibrium off-center Nb displacement from the
cubic phase and the corresponding energy lowering)
has then been done in order to find a more economic setup,
with respect to $L_{max}$ and the basis choice, that
does not result in a noticeably loss of accuracy.

For the K atom with its uncommonly large muffin-tin sphere,
we found it preferable to keep $L_{max}$=6, whereas for
Nb and O atoms with much smaller spheres the cutoff
$L_{max}$=4 suffices, as is consistent with the experience
of other calculations performed with the code by M.Methfessel.
As regards the basis set, we found it necessary to
include all 5$s$, 4$p$ and 4$d$ states combined with
all three Hankel function tails on the Nb site, i.e.
adding $p$ and $d$ states matched with the Hankel function
of the lowest tail energy ($-$2.3 Ry) as compared
with the prescription of Ref.\cite{ktn3}. The choice
of basis functions related to K and O sites
did not need to be changed. Note that the inclusion
of the 3$d$ states on O sites for at least one (the highest)
envelope function energy is essential for describing
the very existence of the ferroelectric instability.
We checked that the further enhancement of the O basis set
using the 3$d$ functions matched to other Hankel envelopes
does not lead to any substantial changes. On the K site,
the basis functions matched with the lowest-lying
Hankel envelope may be thrown out whatsoever.

These refinements do not affect the total energy dependence
on volume in any noticeable way, nor do they result
in substantial changes of the total energy vs. Nb displacement
behavior in cubic or tetragonal phases of KNbO$_3$,
compared to the results of Refs. \cite{ktn3,phonon,wil94}.
However, for the orthorhombic phase characterized by much smaller
energy differences on displacement and tiny magnitudes
of the equilibrium displacements, our present results
may be considered as some improvement over those
preliminarily reported in Ref. \cite{ortho}.

\section{Equilibrium structure of the orthorhobmic phase
	 of KN\lowercase{b}O$_3$ and $\Gamma$-TO phonons}
\label{sec:struc}

\begin{table}
\caption{
Positions of atoms in orthorhombic phase of KNbO$_3$
(in terms of lattice parameters) as determined
by neutron diffraction measurements,
Ref. \protect\cite{hewat}, and optimized
in the FP-LMTO total-energy calculation.
}
\label{tab:equilib}
\begin{tabular}{cllllrldd}
   & Atom & $a$ & $b$ & $c$ & & & ~~~~$\Delta_{exp}$ &
			       ~~~~$\Delta_{calc}$ \\
\hline
 1 & K     & 0~~~~~~ & 0 & $\Delta_z$ & & & 0.0138 & 0.0314 \\
 2 & Nb    & $\frac{1}{2}$ & 0 & $\frac{1}{2}$ \\
 3 & O(I)  & 0 & 0 &
	     $\frac{1}{2}+\Delta_z$ & & & 0.0364 & 0.0351 \\
 4 & O(II) & $\frac{1}{2}$ & $\frac{1}{4}+\Delta_y$ &
	     $\frac{1}{4}+\Delta_z$ &
	     & $\Delta_z$~: & 0.0342 & 0.0353 \\
 5 & O(II) & $\frac{1}{2}$ & $\frac{3}{4}-\Delta_y$ &
	     $\frac{1}{4}+\Delta_z$ &
 \raisebox{2.5ex}[0pt]{$\Bigg\}$} &
	     $\Delta_y$~: & $-$0.0024 & $-$0.0037 \\
\end{tabular}
\end{table}

With the calculation setup as described above, we
performed the total energy-driven optimization of atomic positions
in the orthorhombic phase. In the setting used by Hewat\cite{hewat},
the three lattice vectors are
$a$=3.973{\AA} along {\bf x}=[100] of the cubic aristotype;
$b$=5.695{\AA} along {\bf y}=[0$\bar 1$1], and
$c$=5.721{\AA} along {\bf z}=[011]. The positions of the atoms
have been optimized in a four-dimensional parameter space
as done earlier in Ref.\cite{ortho}, i.e. keeping the
lattice volume and strain constant in agreement
with the experimental data. The displacements of atoms
from their symmetry positions in terms of corresponding lattice
vectors are shown in Table \ref{tab:equilib}. As compared with
earlier FP-LMTO calculations with a slightly different setup\cite{ortho},
the $z$-displacement of K was somehow reduced and
the $z$-displacement of both O(I) and O(II) increased,
resulting in better agreement with the experimental geometry;
the estimate of the small O(II) displacement in the $y$ direction
also became more accurate. Of all structure parameters, only
our estimate of the K $z$-displacement still remains
beyond the experimental uncertainty as given in Ref. \cite{hewat}.

Based on this refined geometry, we calculated the $\Gamma$-TO
phonon frequencies as another benchmark for the quality of our
description of the total energy hypersurface as a function of
atomic displacements. The symmetry coordinates for the
displacements from the equilibrium positions in the $Amm2$
space group of orthorhombic KNbO$_3$ are as given in Ref.\cite{ortho}.
Based on a polynomial fit of the total energy hypersurface
in terms of symmetry coordinates, we calculated the
frequencies of the phonon modes compatible with
the $B_2$ (that including the soft mode)
irreducible representation within our new calculation setup.
The frequencies and eigenvectors are shown in Table \ref{tab:phonon}.
As compared to our earlier calculation \cite{ortho},
we obtained somehow better agreement with the experimental
frequencies. Moreover, the contribution of the K displacement
in the eigenvector of the soft mode decreases,
which means that the soft mode represents primarily
the vibration of Nb against the oxygen sublattice,
with relatively unaffected K in the background.
This seems to be in better agreement with the displacement
pattern of atoms in the rhombohedral structure that emerges
as the $B_2$ soft mode freezes down, and thus fixes
the unaccuracy pointed out in our previous study
where the coupling of K to the oxygen octahedra
was overestimated (see Fig.1 of Ref. \cite{ortho}).
The calculated frequency of the $A_2$ mode is
252 cm$^{-1}$, which is again in somehow better agreement with
the experimental estimate of 283 cm$^{-1}$ \cite{bh76}
than 224 cm$^{-1}$ as calculated in Ref. \cite{ortho}.

\begin{table}
\caption{
Calculated $\Gamma$-TO frequencies and eigenvectors
for the $B_2$ modes in orthorhombic KNbO$_3$.
}
\label{tab:phonon}
\begin{tabular}{dddddcc}
\multicolumn{5}{c}{Eigenvectors (present work)} \\
\cline{1-5}
  K & Nb & O$_1$ & O$_2$ & O$_3$ &
\raisebox{2.5ex}[0pt]{$\omega$ calc. (cm$^{-1}$)} &
\raisebox{2.5ex}[0pt]{$\omega$ exp. (cm$^{-1}$)} \\
\hline
    0.14  & $-$0.62  &    0.63 & 0.32 & 0.32 &~~~~169$i$~~~~&  soft \\
 $-$0.82  &    0.32  &    0.47 & 0.02 & 0.02 & 204 & 195$^a$--197$^c$ \\
 $-$0.30  & $-$0.05  & $-$0.54 & 0.56 & 0.56 & 607 & 511$^a$--516$^c$ \\
\end{tabular}
$^a$ Raman spectroscopy; Ref. \cite{bh76}. \\
$^c$ Infrared spectroscopy; Ref. \cite{fmsg84}.
\end{table}

\section{Chain structures and off-$\Gamma$ frozen phonons}
\label{sec:chain}

A complete mapping of the phonon frequencies over
the Brillouin zone of cubic KNbO$_3$ has been done
by Yu and Krakauer \cite{chain} in an {\em ab initio}
linear response calculation. Straightforward supercell
calculations are too time consuming for any dense mapping
of this kind, nor are they feasible for an arbitrary
{\bf k} point. They may be however useful for going beyond
the harmonic approximation for phonons, which is a natural
limitation of a linear response approach, and for analyzing
the ferroelectric instability in more detail for several selected
{\bf k} points. In principle, the continous mapping
over the Brillouin zone is as well possible based
on a selection of representative supercell calculations
for the fitting of the total energy expansion coefficients,
as has been proposed by Wei and Chou \cite{wei92}.
This is only a question of the computational effort,
but not a principal limitation.

In the present paper, we provide some preliminary
results concerning the chain-structure instability
in orthorhombic KNbO$_3$. We performed the calculations
for the experimental lattice parameters as discussed
in Section \ref{sec:struc}, i.e. for the cell volume very close
to that used in Ref. \cite{chain}. Since the orthorhombic
distortion is rather small, one should not expect
large differences in what regards the region of
ferroelectric instability. We concentrated therefore
on the total energy lowering as a function of finite
atomic displacements, with the ultimate aim to study
the effect of displacement correlations on the
anharmonic stabilization of structure.
For the construction of supercells and the underlying
symmetry analysis, we found the software tools developed
by Stokes \cite{stokes} very useful. When considering
the frozen phonon for a {\bf k} point along a
[100] or [010] symmetry line in an orthorhombic perovskite
structure, the soft mode is mixed among 9 coupled symmetry
coordinates, that demands a corresponding fitting
of the total energy hypersurface. For preliminary estimates,
we considered only [100]-displacement of Nb, that dominate
in the soft-mode eigenvector as is known from
the single-cell calculations. The displacements of
Nb atoms within particular patterns, depending on
the given {\bf k} vector, are from their equilibrium
positions in the orthorhombic structure as discussed
in Section \ref{sec:struc}.

\begin{figure}
\epsfxsize=10.0cm\centerline{\epsffile{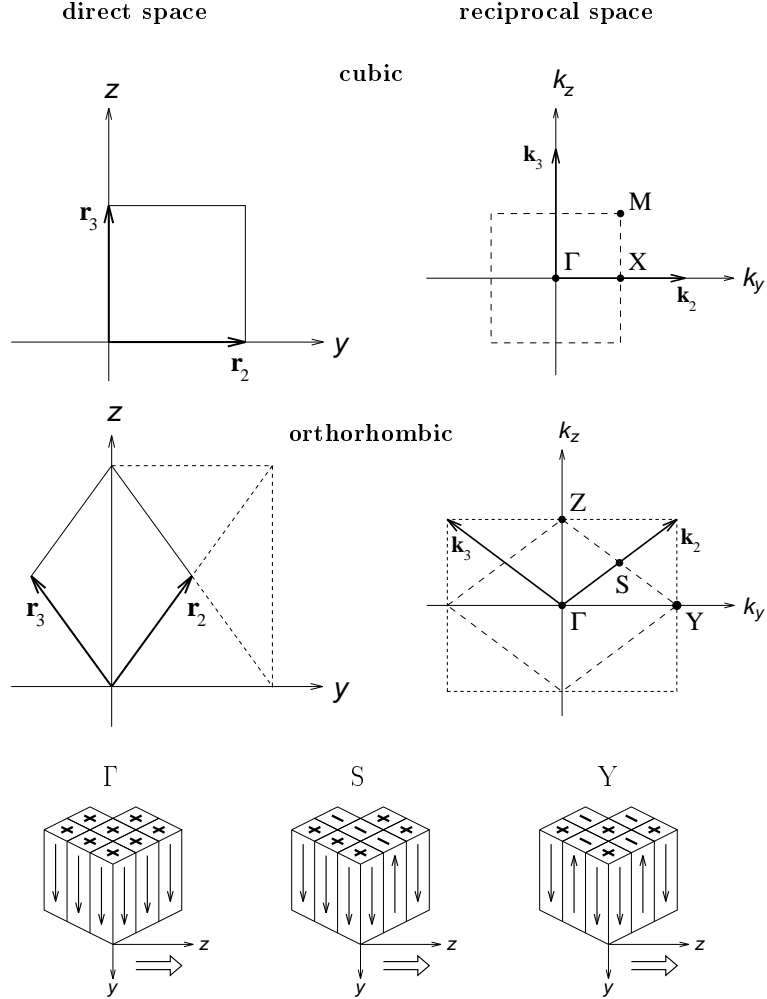}}
\caption{\protect\rule{0mm}{10mm}
Relation between cubic and orthorhombic settings
in direct and reciprocal space (top),
and schematic orientation of dipole moments within chains
for supercells corresponding to $\Gamma$, $S$ and $Y$
{\bf k}-vectors (bottom).
Note that the electric field along chains is
on top of the macroscopic field along the $z$ axis.
}
\label{fig:struc}
\end{figure}

In order to check how the ferroelectric instability
varies over the (100) plane in the reciprocal space,
we considered the points $Y=[010]$ and
$S=[0\frac{1}{2}\frac{1}{2}]$,
that come into $M$ and $X$ of the cubic structure
in the limit of small orthorhombic distortions.
In the direct space, corresponding arrangements are
infinite [100] chains of stockpiled primitive cells,
with Nb atoms displaced along [100] in half of the chains
and in the opposite direction in the other half.
With respect to each other, these ``up'' and ``down'' chains
either pack in the chessboard configuration ($Y$),
or stick to form interchanging
planes ($S$). The relations between cubic and
orthorhombic settings and the relevant real-space
superstructures are shown in Fig.\ref{fig:struc}.

\begin{figure}
\epsfxsize=14.5cm\centerline{\epsffile{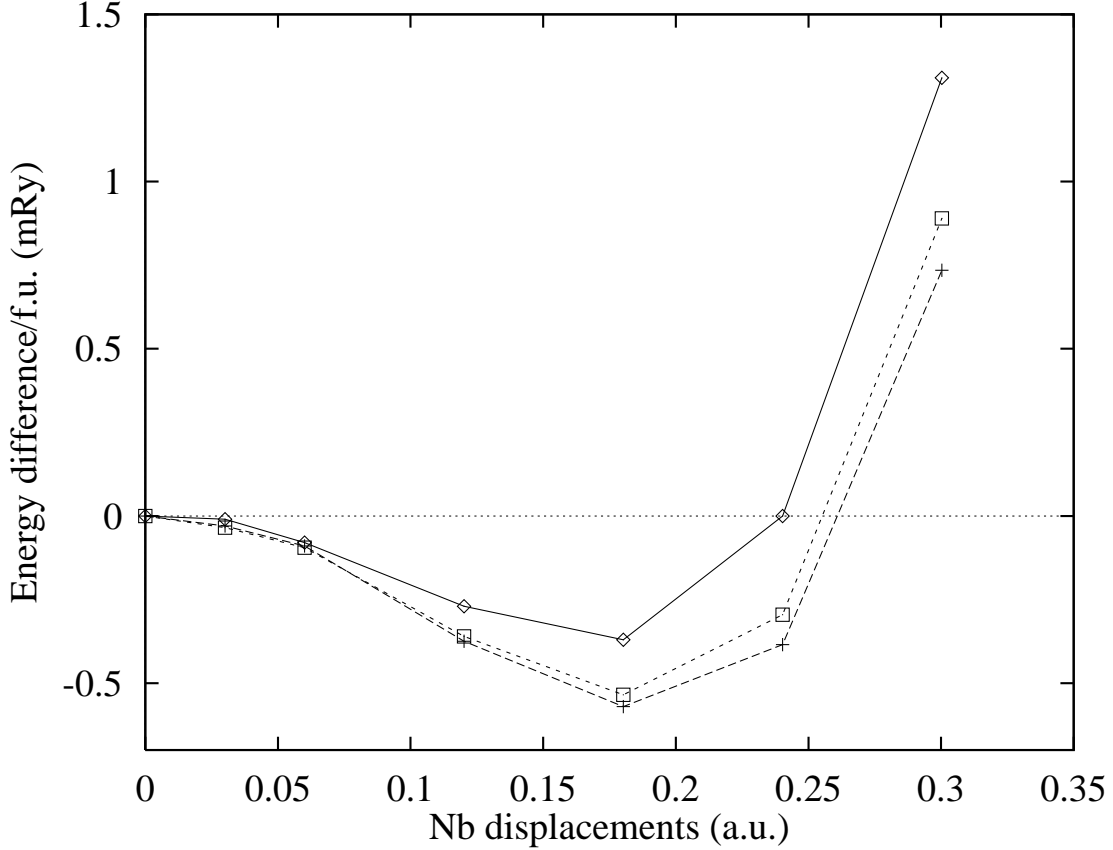}}
\caption{\protect\rule{0mm}{10mm}
Total energy difference (per formula unit)
as a function of the [100] Nb displacement
for the TO $\Gamma$, $S$ and $Y$ frozen phonons.
}
\label{fig:deltae}
\end{figure}

One can expect from the analysis of Ref. \cite{chain}
that the force constant related to the soft mode
is not changed significantly along the $\Gamma$--$X$--$M$
directions. It follows from our earlier study
of different Nb displacement patterns \cite{ortho}
that the force constants corresponding to $\Gamma$
and $Y$ are indeed very close (see Fig.3, patterns
$a$ and $b$, of Ref. \cite{ortho}). However, the
chessboard confguration of chains turns out to be
more stable, allowing larger magnitude of Nb
displacements within the chains.
The present study supports this conclusion,
now with the magnitudes of displacements and
the energy gains somehow adjusted with our refined
calculation setup, and the data for the $S$ frozen
phonon added. The total energy gain as a function of
Nb displacements is shown in Fig. \ref{fig:deltae}.
As may be intuitively expected, the curve corresponding
to the $S$ phonon is the intermediate one between those
for $\Gamma$ and $Y$, but rather close to the latter.
This type of mutual arrangement can be
easily understood from electrostatic considerations.
Electric dipoles, created (on top of the [001] macroscopic
field) by additional [100] or [$\bar 1$00] Nb displacements
in individual orthorhombic cells, tend to form chains,
but dipoles from different chains, when the latter
come closer, have an energy gain if oriented
antiparallel. The chessboard configuration
realizes the maximal separation between parallel
dipoles on a (100) two-dimensional lattice.
Due to the anisotropy of the dipole field,
the dipole-dipole interaction between chains is
much weaker than inside a chain. As a consequence, the
ideal chessboard arrangement is not bound to occur
at temperatures of the order 300 K $\sim$1.9 mRy;
instead, the chains may be considered as relatively
independent (and irregular) in their orientation,
as long as the number of ``up'' and ``down'' chains
remains roughly equal\cite{coment}.

\begin{figure}
\epsfxsize=8.5cm\centerline{\epsffile{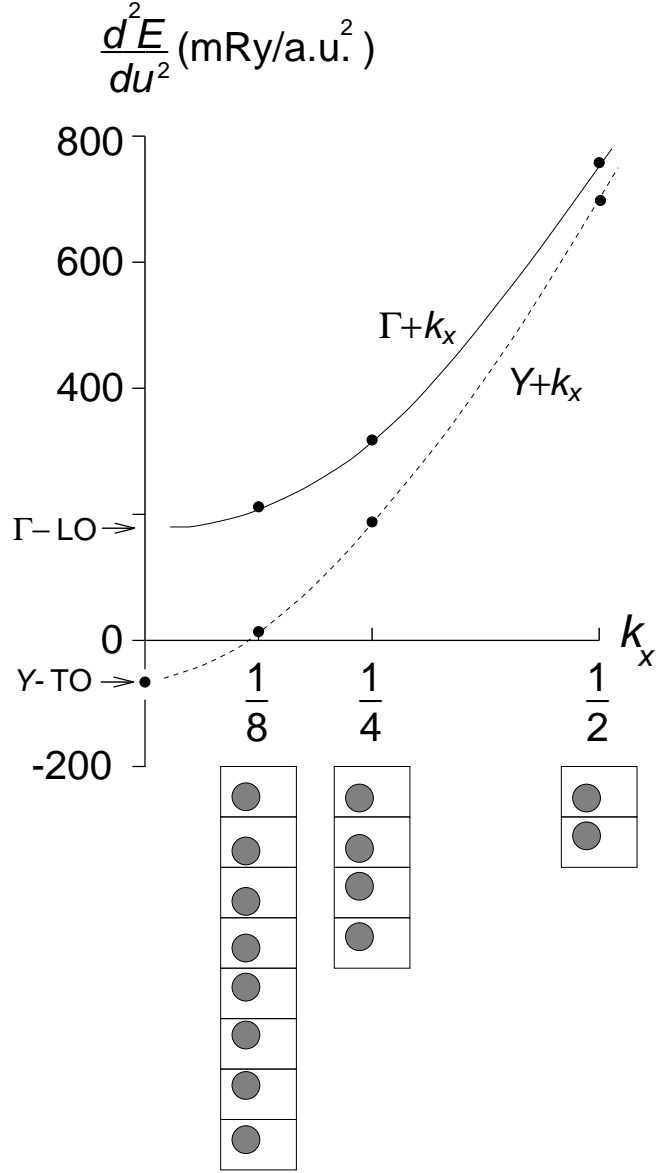}}
\caption{\protect\rule{0mm}{10mm}
Force constant of the [100] Nb displacement
for TO $[0\,1\,k_x]$ (dots on a dashed line)
and LO $[0\,0\,k_x]$ (dots on a solid line) phonons
as determined from supercell calculations.
The supercells used and corresponding Nb displacement patterns
are shown below.
}
\label{fig:LO}
\end{figure}

This consideration effectively sets the correlation
length within the (100) plane, i.e. in the direction
normal to chains, to one lattice parameter.
The analysis of the correlation
length along chains demands to go away from the (100) plane
in the reciprocal space, e.g., along $M{\rightarrow}R$
in the cubic setting, and considers the hardening
condition of the corresponding transversal phonon mode.
Based on a $\omega^2=0$ condition for a harmonic frequency
at a critical correlation length, Yu and Krakauer \cite{chain}
set the latter at about 5 lattice spacings. For the orthorhombic
lattice, we did the corresponding analysis with a
frozen phonon calculation for several ${\bf q}$ points
on a $Y{\rightarrow}T$ line.
The values of the force constant related to the [100] displacement
of Nb as calculated for several increasingly larger supercells
of this type are shown in Fig.\ref{fig:LO} and connected
(as a guideline) by a dashed curve labelled $Y+k_x$.
The supercells have the arrangement of chains
as is shown in Fig.\ref{fig:struc} in the (100) plane;
the [100] Nb-displacements within the chains
are however now varied as $\Delta_x$(Nb)$=u\cos(k_x x)$.
Each supercell represents therefore a body-centered orthorhombic
structure with 2, 4 or 8 formula units per primitive cell;
the force constant is the total energy per unit cell
twice differentiated in the displacement amplitude $u$.
For $k_x=0$, the value $-$65.6 mRy/a.u.$^2$
obtained from the double-cell calculation of the $Y$ phonon
(as shown in Fig.\ref{fig:deltae}) is plotted.
The sketches in Fig.\ref{fig:LO} below
show the [100] Nb displacement pattern in the supercells.
One can see that the ferroelectric instability develops
at $k_x\le0.11$, i.e. the critical half-period
of the Nb displacement wave that creates
a self-supporting instability is between
4 and 5 unit cells. This is in agreement with
the estimate done by Yu and Krakauer\cite{chain}.
For a more accurate analysis however, one should construct and
diagonalize the 9$\times$9 dynamicial matrix and search
for the condition of a zero eigenvalue to occur.

As another set of examples for off-$\Gamma$ frozen phonons,
we considered several $\Lambda$ points on a [100] line,
that corresponds to the uniform Nb displacement pattern
over the (100) plane, but with harmonic variation
(represented by increasingly larger supercells)
of $\Delta_x$(Nb)$\sim\cos(k_x x)$ in the normal direction.
The values of the force constant corresponding
to this setting are shown in Fig.\ref{fig:LO} as laying
at a solid line labelled $\Gamma+k_x$.
Since in this case
the phonon mode in question is longitudinal, it should not
be expected to get soft even at small values of $k_x$.
The problem of interest in this case is the extrapolation
of the force constants towards those for the $\Gamma$-LO
phonon. The conventional way of treating the $\Gamma$-LO
phonons is via correcting the dynamical matrix to
Born effective charges (see, e.g., \cite{loto,resta}).
Although physically transparent, this method incorporates
parameters which have to be extracted from different schemes.
Born effective charges are usually fitted from
calculations of the finite bulk polarization;
the optical macroscopic dielectric function
$\varepsilon_{\infty}({\bf q},\omega)$ is either extrapolated
to $\omega=0$ from experimental data (Zhong {\it et al.}
cite the value $\varepsilon_{\infty}$=4.69
for KNbO$_3$\cite{loto}), or obtained from a linear
response calculation, with a typically overestimated,
due to the local density approximation, value
(Yu and Krakauer report $\varepsilon_{\infty}$=6.34 \cite{chain}).
The supercell calculation for the LO phonons, on the contrary,
involves only the total energy of the ground state
(in principle, without any limitation as regards the anharmonicity),
and the effects of long-range polarization and dielectric response
are already implicitly included.
Taking into account that the force constants
should be symmetric with respect to $k_x \rightarrow -k_x$,
we extrapolated the force constant related to the [100] Nb displacement
in the $\Gamma$-LO mode to be 177.2 mRy/a.u.$^2$ (see Fig.\ref{fig:LO}).
The corresponding force constant in the $\Gamma$-TO mode
(as extracted from a single-cell calculation;
see Fig.\ref{fig:deltae}) is $-$52.3 mRy/a.u.$^2$.
Based on the relation between LO and TO force constants
(see Ref.\cite{pick}, or the relevant formulas (2) and (3)
in Ref.\cite{loto}), we arrive at the following relation
between the $xx$-component of the Nb dynamic effective charge
$Z^{*}_{Nb}$ and $\varepsilon_{\infty}$:
$(Z^{*}_{Nb})^2/{\varepsilon_\infty}=3.94$.
For the two above cited choices of $\varepsilon_{\infty}$,
this leads to an estimate of the relevant
Born effective charge between 4.3 ($\varepsilon_{\infty}$=4.69)
and 5.0 ($\varepsilon_{\infty}$=6.34).
One should note that the Nb-related effective charge tensor
is no more diagonal in the orthorhombic structure,
and we make no estimates for other elements of it
than $Z^{*}_{xx}$ at the moment.
Generally, the value of the $xx$ element in the orthorhombic phase
may be expected to be smaller than
9.13 \cite{resta} to 9.37 \cite{chain} as calculated
for the cubic perovskite structure by different methods,
because the off-center displacement of Nb from the center
of a cubic perovskite cell to its position in the orthorhombic
cell (along [001], in the present setting)
was already accompanied by some charge transfer that reduces
the polarizability of the Nb--O bonds in perpendicular directions.
Moreover, the [100] displacement of Nb in the orthorhombic phase
is not exactly toward an oxygen atom, as it was the case
in the cubic structure.

\section{Conclusion}
\label{sec:conclu}

Based on our previous experience in total-energy calculations
with the full-potential LMTO code by M.~Methfessel
and on the systematic study of the effects of
basis choice and the spherical harmonic cutoffs,
we attempted to achieve the highest degree of accuracy,
attainable with the present status of the code,
in the description of microscopic structure
of ferroelectric KNbO$_3$ in the room-temperature
orthorhombic phase.

This resulted in somehow more accurate, as compared
to our previous data, description of the equilibrium
geometry, i.e. of optimized off-center displacements of
atoms, with the lattice parameters and the orthorhombic strain
kept fixed. The $\Gamma$-TO phonon frequencies
calculated for the $B_2$ mode are on average
in the same degree of agreement with experimental data
as in earlier calculations, and the frequency for the $A_2$ mode
has improved. Making use of the computational
efficiency of the code, we attempted
to model several more complicated frozen phonon patterns
in the supercell calculations with up to 40 atoms,
aiming at the study of the chain instability and
of long-wavelength LO modes. The correlation length for the
onset of the ferroelectric instability is estimated to be
between 4 and 5 lattice constants in the [100] direction,
i.e. along the chains of displaced Nb atoms, which is consistent
with the results of a linear response calculation by
Yu and Krakauer. The correlations between different chains
(in the perpendicular direction) are found to be
unimportant for the onset of the ferroelectric instability,
but they stabilize the direction of macroscopic polarization
compatible with the orthorhombic symmetry, as long as
the orthorhombic lattice strain is fixed.

The comparison of force constants, as obtained directly
for the $\Gamma$-TO mode and extrapolated to ${\bf k}\rightarrow 0$
from supercell calculations for a LO mode, makes it possible
to exploit the relation between Born effective charges
and the optical dielectric constant as a tool to estimate
whichever of these properties is unknown for the system in question.

\acknowledgements

The authors are grateful to M.~Methfessel for his assistance and
advise in using his full-potential LMTO code, and to H.~Stokes
for providing the symmetry-analysis software.
Financial support of the Deutsche Forschungsgemeinschaft (SFB~225)
is gratefully acknowledged.


\begin{references}
\bibitem[*]{*}   Electronic address: apostnik@physik.uni-osnabrueck.de
\bibitem{fmsg84} M.~D.~Fontana, G.~M\'etrat, J.~L.~Servoin,
		 and F.~Gervais,
		 {\it J.~Phys.~C}, {\bf 17}, 483 (1984).
\bibitem{frkc88} M.~D.~Fontana, A.~Ridah, G.~E.~Kugel, and
		 C.~Carabatos-Nedelec,
		 {\it J.~Phys.~C}, {\bf 21}, 5853 (1988).
\bibitem{scr88}  J.~P.~Sokoloff, L.~L.~Chase, and D.~Rytz,
		 {\it Phys.~Rev.~B}, {\bf 38}, 597 (1988).
\bibitem{fkc89}  M.~D.~Fontana, G.~E.~Kugel, and C.~Carabatos-Nedelec,
		 {\it Phys.~Rev.~B}, {\bf 40}, 786 (1989).
\bibitem{scr89}  J.~P.~Sokoloff, L.~L.~Chase, and D.~Rytz,
		 {\it Phys.~Rev.~B}, {\bf 40}, 788 (1989).
\bibitem{js62}   F.~Jona and G.~Shirane,
		 ``Ferroelectric Crystals,''
		 (Pergamon, New York, 1962).
\bibitem{fkmc81} M.~D.~Fontana, G.~E.~Kugel, G.~Metrat, and C.~Carabatos,
		 {\it Phys. Status Solidi (b)}, {\bf 103}, 211 (1981).
\bibitem{dwng92} T.~P.~Dougherty, G.~P.~Wiederrecht, K.~A.~Nelson,
		 M.~H.~Garret, H.~P.~Jensen, and C.~Warde,
		 {\it Science}, {\bf 258}, 770 (1992).
\bibitem{doug1}  T.~P.~Dougherty, G.~P.~Wiederrecht,
		 and K.~A.~Nelson,
		 {\it Ferroelectrics}, {\bf 164}, 253 (1995).
\bibitem{doug2}  T.~P.~Dougherty, G.~P.~Wiederrecht, K.~A.~Nelson,
		 M.~H.~Garrett, H.~P.~Jenssen, and C.~Warde,
		 {\it Phys.~Rev.~B}, {\bf 50}, 8996 (1994).
\bibitem{wil94}  A.~V.~Postnikov, T.~Neumann, and G.~Borstel,
		 {\it Ferroelectrics}, {\bf 164}, 101 (1995).
\bibitem{sb92}   D.~J.~Singh and L.~L.~Boyer,
		 {\it Ferroelectrics}, {\bf136}, 95 (1992).
\bibitem{ktn3}   A.~V.~Postnikov, T.~Neumann, G.~Borstel and
		 M.~Methfessel,
		 {\it Phys.~Rev.~B}, {\bf 48}, 5910 (1993).
\bibitem{ksv94}  R.~D.~King-Smith and D.~Vanderbilt,
		 {\it Phys.~Rev.~B}, {\bf 49}, 5828 (1994).
\bibitem{phonon} A.~V.~Postnikov, T.~Neumann, and G.~Borstel,
		 {\it Phys.~Rev.~B}, {\bf 50}, 758 (1994).
\bibitem{ortho}  A.~V.~Postnikov and G.~Borstel,
		 {\it Phys.~Rev.~B}, {\bf 50}, 16403 (1994).
\bibitem{loto}   W.~Zhong, R.~D.~King-Smith, and D.~Vanderbilt,
		 {\it Phys.~Rev.~Lett.}, {\bf 72}, 3618 (1994).
\bibitem{comes}  R.~Comes, M.~Lambert, and A.~Guinier,
		 {\it Solid State Commun.}, {\bf 6}, 715 (1968).
\bibitem{chen}   M.~Holma, N.~Takesue, and H.~Chen,
		 {\it Ferroelectrics}, {\bf 164}, 237 (1995).
\bibitem{chain}  Rici Yu and H.~Krakauer,
		 {\it Phys.~Rev.~Lett.}, {\bf 74}, 4067 (1995).
\bibitem{msm1}   M.~Methfessel,
		 {\it Phys.~Rev.~B}, {\bf 38}, 1537 (1988).
\bibitem{msm2}   M.~Methfessel, C.~O.~Rodriguez, and O.~K.~Andersen,
		 {\it Phys.~Rev.~B}, {\bf 40}, 2009 (1989).
\bibitem{hewat}  A.~W.~Hewat,
		 {\it J.~Phys.~C}, {\bf 6}, 2559 (1973).
\bibitem{bh76}   D.~G.~Bozinis and J.~P.~Hurrell,
		 {\it Phys.~Rev.~B}, {\bf 13}, 3109 (1976).
\bibitem{wei92}  S.~Wei and M.~Y.~Chou,
		 {\it Phys.~Rev.~Lett.}, {\bf 69}, 2799 (1992).
\bibitem{stokes} H.~T.~Stokes,
		 {\it Ferroelectrics}, {\bf 164}, 183 (1995).
\bibitem{coment} Obviously enough, the rhombohedral phase
		 with all chains, say, in ``up'' configuration
		 must have lower total energy as a result
		 of optimization in volume along with
		 {\it all} coordinate and strain variables.
		 The structure compatible with the $Y$ phonon
		 comes out as the most stable one in our
		 calculation under the constraint of fixed orthorhombic
		 lattice parameters.
\bibitem{resta}  R.~Resta, M.~Posternak, and A.~Baldereschi,
		 {\it Phys.~Rev.~Lett.}, {\bf 70}, 1010 (1993).
\bibitem{pick}   R.~M.~Pick, M.~H.~Cohen, and R.~M.~Martin,
		 {\it Phys.~Rev.~B}, {\bf 1}, 910 (1970).
\end{references}
\end{document}